# Electrical Charge Control of *h*-BN Single Photon Sources

**Mihyang Yu[1,2], Donggyu Yim[2], Hosung Seo[2] and Jieun Lee[1,2]**

[1] Institute of Applied Physics and Department of Physics and Astronomy, Seoul National University, Seoul, 08826, Korea
[2] Department of Physics and Department of Energy Systems Research, Ajou University, Suwon, 16499, Korea

E-mail: lee.jieun@snu.ac.kr

## Abstract

Colour centres of hexagonal boron nitride (*h*-BN) have been discovered as promising and practical single photon sources due to their high brightness and narrow spectral linewidth at room-temperature. In order to realize *h*-BN based photonic quantum communications, the ability to electrically activate the single photon fluorescence using an external electric field is crucial. In this work, we show the electrical switching of the photoluminescence from *h*-BN quantum emitters, enabled by the controllable electron transfer from the nearby charge reservoir. By tuning the Fermi level of graphene next to the *h*-BN defects, we observed luminescence brightening of a quantum emitter upon the application of a voltage due to the direct charge state manipulation. In addition, the correlation measurement of the single photon sources with the graphene's Raman spectroscopy allows us to extract the exact charge transition level of quantum emitters, providing the information on the crystallographic nature of the defect structure. With the complete on-off switching of emission intensity of *h*-BN quantum emitters using a voltage, our result paves the way for the van der Waals colour centre based photonic quantum information processing, cryptography and memory applications.

## 1. Introduction

Hexagonal boron nitride (*h*-BN) is a two-dimensional (2D) crystal, which has the great potential for the development of novel nanoscale devices [1–3] because of its excellent optical, electrical, thermal and mechanical properties. Due to its ultra-flat surface, high thermal stability and wide band gap (~ 6 eV), *h*-BN has been useful as ingredients for van der Waals devices such as quantum capacitors, field effect and tunneling transistors, and various quantum optoelectronic devices [4–7]. More recently, *h*-BN has been found to host defect-derived single photon sources that can generate stable and bright single photon even at room temperature, introducing *h*-BN as an emerging material platform for the solid-state quantum information technologies [8,9]. Various progresses have been made to study the nature and controllability of the quantum emitters including the identification of the emitter's crystallographic structure through theoretical modeling [10–15], characterization of their optical properties using various experimental tools [16–19], and the manipulation of the emission properties through external stimuli such as strain and electric field [20–22]. Moreover, the artificial device geometries were developed by employing the unique planar nature of *h*-BN such as the localized creation of quantum emitters on prepatterned substrates [23] and coupling the emitters with nano-photonic structures [24].

For the exploitation of *h*-BN quantum emitters for quantum communication applications, however, the full electrical manipulation of quantum emitters is highly desired. Although the electrical control of the emitter's photon energy was reported in several experiments by applying electric fields to *h*-BN defects [21,25,26], the direct manipulation of the charge state of *h*-BN quantum emitters through a voltage has not been achieved yet. By using a systematic charge transfer into *h*-BN quantum defects, it will become possible to induce electrically activated quantum emitters through doping [27–29].

Moreover, *h*-BN quantum emitters are expected to have different emission energy, spin configuration and brightness depending on the defect's charge state [21,28–33]. Up to date, many theoretical works were reported to figure out the emitter's crystallographic structures and associated charge states. For example, one of the widely studied defect candidates in *h*-BN is that consisting of a nitrogen vacancy centre next to a nitrogen antisite ($V_NN_B$) which is proposed to have positive, neutral and negative charge states within the band gap with strong zero phonon line (ZPL) at 2 eV for $V_NN_B^0$ [11]. In addition, other defect candidates are also proposed such as a negatively charged boron vacancy ($V_B^-$), a neutral complex defect composed of a nitrogen vacancy centre next to a substitutional carbon ($V_NC_B^0$), and a negatively charged boron vacancy with dangling bonds (BVDB$^-$) [11,13,14]. However, in spite of many theoretical proposals on the charge states of *h*-BN quantum emitters, experimental studies on the crystallographic structure of defect emitters and their associated charge states are still unknown.

In this work, we use graphene as a charge reservoir with the largely tunable Fermi level [34,35] to induce the photoluminescence switching of *h*-BN local defects through the electrical charge state manipulation. The electrical onset of the *h*-BN defect emission is induced by the Fermi level tuning of graphene layers, allowing the extraction of the exact charge transition level of *h*-BN defects, providing information on the possible defect model candidate with the known charge state. Also, we observed the voltage-induced peak splitting of a quantum emitter, which may have implications for the coherent coupling of different states using quantum light sources. Our work demonstrating the electrical activation of *h*-BN quantum emission will not only be advantageous for the crystallographic identification of *h*-BN defects but also for the development of 2D optoelectronic single photon devices and van der Waals based quantum communications. By achieving the controllable manipulation of the charge state in single photon sources in *h*-BN, further applications in defect's quantum controls will also become possible [30,36,37].

## 2. Results

### *2.1 Device and electrical characterization*

The schematic of the electrical control of the *h*-BN quantum emission in our work is shown in figure 1(a). The atomic defects in *h*-BN crystals are first prepared by thermal annealing of exfoliated *h*-BN (~ 20 nm thick). To induce the systematic charge transfer from graphene to *h*-BN defects, we fabricated heterostructures consisting of graphene/*h*-BN/graphene stacked layers (microscopic image and device schematic in figure 1(b) and figure 2(a), respectively). With the fabricated device, we performed electrical test measurements of graphene electrodes (See supplementary figure S1). From the detailed measurement of the $I-V$ curve of our device, the Fowler-Nordheim tunnelling is observed, which is caused by the electric field induced Fermi level tilting and charge tunnelling, confirming the effective charge doping of quantum defects in *h*-BN crystals [38–40].

The Fermi level tuning of the graphene electrode could also be independently probed by the Raman spectroscopy as shown in figure 2. By scanning the voltage across graphene in the range from -9 to 9 V, the graphene's Fermi level energy was tuned by about 500 meV as verified from the shift of the graphene Raman G peak. The detailed peak shift and linewidth narrowing of the graphene G peak are shown in figure 2(c). The characteristic Kohn anomaly induced Raman shift of graphene near $E_F \approx \pm 100$ meV is found [41], indicating the clean interface of the fabricated heterostructure and the effective Fermi level tuning of graphene layers. The measured range of Fermi level tuning also agrees well with the calculation using the geometric and quantum capacitance of graphene electrodes (See Methods).

*2.2 Electrically activated emitter photoluminescence*

With information of the graphene's Fermi level control, we performed 2D spatial scanning of the photoluminescence from *h*-BN devices while varying the applied voltage ($V$) as shown in figure 1(c and d). All measurements were performed at cryogenic temperature at about 10 K. A continuous-wave 532 nm laser was used as an excitation source and we spatially scanned the beam position over the heterostructure to obtain 2D photoluminescence map. When we applied zero voltage to the device ($V = 0$), no luminescence of light was observed in the area that we scanned as shown in figure 1(c). However, when we turned on the external voltage ($V = -6$ V) and scanned over the same area, several localized luminescence of light was observed at different locations. To better display this, we plotted the wavelength dependent 2D photoluminescence map of *h*-BN in figure 1(d). The microscope image of the scanned area is indicated by a box in figure 1(b). In this area (3 µm × 5 µm), we found 4 electrically activated *h*-BN emitters which have emission wavelength ranging from 580 to 620 nm. Interestingly, these voltage-induced *h*-BN emitters showed markedly different voltage dependence between emitters which could be categorized into two kinds, *i.e.,* one with asymmetric voltage dependence and another with symmetric voltage dependence, suggesting the existence of different activation mechanisms, which are described below.

*2.3 Quantum emitters with asymmetric voltage dependence*

Figure 3(a) shows the voltage dependent spectra from a localized emission shown in figure 1(d), which is denoted as emitter E1. In this emitter, the luminescence is switched on at a threshold voltage of about $-4$ V. The emission spectrum at positive, zero, and negative applied voltages are plotted with vertical shifts in figure 3(b). The emitter was found to be luminescent only at negative applied voltages. The emitted light is found to have a linear polarization (supplementary figure S2). Also, as we performed the second order correlation function measurement at $V = -9$ V, the emitter exhibited $g^2(0)$ value below 0.5, confirming that the emitter is a single photon source (figure 3(c)). When we sequentially applied the external voltage of $-9$ V and 0 V repeatedly to the device with 1 s time interval (figure 3(c)), the emitter was turned on and off repeatedly, revealing the excellent switching repeatability. The emitter was stable even after we applied more than 3000 cycles of external on/off voltages (spectrum before and after the switching sequences are shown in supplementary figure S3). In our experiment, similar emitters with asymmetric voltage dependence were frequently observed, but they showed a range of threshold voltages centred around 3.5 V which could be either positive or negative. Some examples are shown in supplementary figure S4.

The observation of the asymmetric voltage dependence of the electrically activated quantum emitters suggests that the activation is originated from the charge transfer process between the graphene reservoir and *h*-BN defects, altering the emitter's charge state. By controllably doping the graphene charge reservoir, the emitter's luminescence properties change when the graphene's Fermi level crosses the charge transition level (CTL) of the *h*-BN defect. Therefore, this charge transition induced emitter switching depends on the applied voltage asymmetrically. By taking the band diagram of graphene and *h*-BN into account (figure 3(e and f)), we find that the measured CTL energy of the emitter E1 is about $4.0 - 4.8$ eV above the valence band maximum of *h*-BN as estimated from the emitter luminescence threshold voltage.

The value is obtained by fully considering the field-induced modification of the band structure (See supplementary figure S5 and S6), the exact value of the Fermi level tuning of graphene (upper axis of figure 2(c)), and the depth of quantum emitters embedded beneath the surface of *h*-BN which has the typical range of about 0 – 3 nm [42]. Due to the thermal annealing process, the quantum emitters in *h*-BN are more frequently found near the surface of *h*-BN which is close to the top graphene.

Furthermore, above experimental results lead us to identify the crystallographic nature of the observed quantum emitter. Among various defect model candidates in *h*-BN crystal [11,13,14], we locate the boron vacancy with dangling bonds (BVDB) [14,43] as the most probable single photon source which agrees well with our experiment. The negatively charged BVDB emitter (BVDB$^-$) is calculated to have a bright zero phonon line (ZPL) emission around 2 eV with two-fold absorption and emission polarization angle dependence. The neutral BVDB emitter (BVDB$^0$), on the other hand, has ZPL energy at 3.27 eV but this emission is weak due to the significant electron-phonon coupling [14]. The calculated CTL energy between BVDB$^-$ and BVDB$^0$ is at 4.8 eV above the valence band maximum of the *h*-BN [14], which is in excellent agreement with our experimental results. From the measured wavelength of the observed emitters, many of the asymmetrically voltage activated emitters have energy close to 2 eV which further supports BVDB$^-$ as a highly possible emitter model of our observation.

Nevertheless, there are also other candidate emitter models that cannot be ruled out for the origins of the experimental observations. These include $V_N N_B$ [11], $V_B$ [11], $V_N C_B$ [13,44], and $C_2 C_N$ [45,46]. We have summarized their calculated ZPL energy and CTL in the table T1 in the supplementary. Since these emitter models also have ZPL energy around 2 eV with linearly polarized emission and moderate CTL energy approachable within the experimental range, they could also be the origin of the other voltage activated emitters in *h*-BN. For example, for the emitters activated only at positive voltages, these other emitter models may provide their crystallographic origins with possible charge states.

## *2.4 Quantum emitters with symmetric voltage dependence*

On the other hand, we also observed another type of voltage-activated emitters that showed symmetrical dependence on the external voltage. Figure 4(a) shows the electrically activated light emission from a *h*-BN defect centre which exhibits the emission onset at both positive and negative applied voltages. The detailed measurement spectra are shown in figure 4(b). By applying an external voltage, the emission luminescence emerges, revealing the emitter that was otherwise not observable. The voltage-activated emitter also shows the electric field induced linear Stark shift (0.19 meV per MV/cm), corresponding to the out-of-plane dipole moment of 0.09 Debye. In addition, the second order correlation function measurement at $V = 7$ V supports that the emitter is a single photon source (figure 3(c)). Notably, this single photon emitter shows the symmetric voltage dependence of the emission onset, suggesting that the voltage activation is not originating from the defect's charge state control since the oppositely charged defects would produce emission with different energies [14]. Rather, the symmetric activation of the single photon source could be driven by the doping-induced enhancement of the emission efficiency.

The doping-induced emission enhancement can be understood by the saturation effect of the charge trap sites commonly found around solid-state quantum emitters. These charge traps could be consisting of nearby local defects or metastable excited states [19,47] which contribute to the reduction of the radiative emission process. By tuning the Fermi level of graphene, the saturation of the nearby trap sites is induced through the charge doping from the graphene reservoir, bypassing the non-radiative emission channel and enhancing the emission intensity. Such voltage-induced trap site saturation occurs similarly for both positive and negative applied voltages due to the random distribution of the charge trap sites. Alternatively, the observed phenomena may also be explained by the photo-doping effect [48–50], but we exclude this possibility since the voltage thresholds for the onset of the luminescence was independent of the pump laser power as observed from another similar emitter (See supplementary figure S7 and S8).

Before we move on, we discuss the statistical results of the voltage activated emitters found in our devices. By performing a larger area photoluminescence scanning measurement to cover the entire device area shown in figure 1(b), we found total 175 emitters. Among them, 124 emitters were observable under zero bias voltage which constitutes the majority of the observed emitters. By applying a voltage, we found 51 additional emitters that were electrically activated. 17 and 34 of them showed asymmetric and symmetric voltage dependence, respectively. Examples of other symmetrically activated emitters are shown in supplementary figure S9. We also note that the emission energy and the slope of the Stark shift in the symmetrically activated emitters varied in the range 1.6 – 2.2 eV and 0.029 – 0.86 meV per MV/cm, respectively, suggesting the possible diverse origins of the emitter structures.

*2.5 Voltage-induced quantum emitter peak splitting*

Lastly, we report an emitter that we observed which exhibits the intriguing voltage-induced emitter peak splitting. Figure 5(a) is the voltage scan result of the emitter, which shows an additional peak at 564 nm appearing at $V = 5$ V in the presence of an original peak at 560 nm. The emission polarizations of the two peaks are aligned to the similar axis as displayed in figure 5(d). Since the emerging peak has the similar polarization with the original peak and the two peaks displayed simultaneous spectral jumps (See supplementary figure S10), we confirm that both peaks originate from the same emitter. The energy difference between the original peak and the emerging peak was about 15 meV and the magnitude of the extracted electric dipole moment of each peak was measured to be 0.065 and 0.26 Debye, respectively. Since the dipole moment of the emerging peak matches well with the previous result [21], this emerging peak is likely to be generated due to the field-induced out-of-plane dipole moment. By checking the emission stability at zero voltage, we confirmed that the small electric dipole moment measured for the original peak is indeed induced by an electric field (supplementary figure S11). Another emitter with similar properties is also shown in supplementary figure S12.

For the emergence of the new peak in this emitter, the voltage-induced development of the second potential well in the emitter's excited-state adiabatic potential energy surface (APES) could be an origin as shown in figure 5(e). In the absence of an external voltage, the excited-state has a single-well APES, but it has a meta-stable point close in the configurational coordinate, which is associated with an atomic tilt in the out-of-plane direction, providing extra degrees of freedom to the defects. Upon the application of a voltage, the defect can lower the energy of the meta-stable configuration owing to the electric-field induced energy shift, associated with the emerging dipole in the out-of-plane direction. As a result, the second ZPL luminescence can be generated through the modified APES. The model is in good agreement with experimental observations since the observed peak splitting energy in experiment is about a few tens of meV. Also, the measured dipole moment of the second ZPL has a significantly larger value than the original one in consistent with the field-induced structural modifications. Moreover, reversing the direction of the electric field will raise the energy of the second potential well, resulting in the asymmetric voltage dependence of the emerging peak. Our observation of the electric-field induced level splitting in a single photon source requires further studies to better understand its characteristics, which may lead to the demonstration of the coherent coupling of different states in quantum emitters and the entangled photon pair generation in *h*-BN crystals [51].

## 3. Conclusions

To conclude, we observed electrically induced photon emission switching and level splitting in *h*-BN quantum emitters by controlling the Fermi level tuning of nearby graphene. The onset of the emitter photoluminescence beyond the threshold voltage is found under unidirectional and bidirectional applied voltages which originate from the direct charge state manipulation of the defect centre such as BVDB$^-$ ↔ BVDB$^0$ and the electrical saturation of the nearby charge trap sites, respectively. The voltage-induced activation of the peak splitting was also observed which is explained by the formation of the second potential well in the excited-state potential energy surfaces. Such peak splitting is also expected to be possible using an in-plane electric field scheme or by breaking the symmetry of defect structures. Also, our results on remarkable on/off control of emitter photoluminescence using a voltage will be useful for the development of optoelectronic single photon devices, opening up a route towards 2D chip-based solid-state quantum information processing and communication. We also note that we recently came across a paper reporting similar results in reference [52].

## Methods

### Sample preparation

*h*-BN crystal (HQ graphene) was mechanically exfoliated onto SiO$_2$/Si substrates by using scotch tapes to obtain thin flakes. Exfoliated *h*-BN flakes were rinsed in acetone for 10 mins and in isopropyl alcohol (IPA) for 5 mins to remove tape residues. Then *h*-BN flakes were thermally annealed for 2 h at 800 ˚C under the condition of argon gas flow (1 Torr) to create optically active *h*-BN defects. For the preparation of graphene electrodes, monolayer and bilayer graphene were obtained through mechanical exfoliation from graphite. We used a dry transfer method using polycarbonate (PC) film to fabricate graphene/*h*-BN/graphene stacks which are transferred directly onto SiO$_2$/Si substrates with prepatterned Au electrodes. PC residue was then sequentially removed by dipping the sample in chloroform for 8 h, acetone for 30 mins, and IPA for 30 mins.

### Optical measurements

In the optical confocal microscopy setup, a 532 nm solid-state continuous-wave (cw) laser was incident onto the sample through an objective lens (NA = 0.6). The emitted light was collected using the same objective and reflected at a beam splitter and guided to a spectrometer with CCD detector. XY-piezostages inside a cryostat were used to move the sample position to perform 2D spatial scanning. The excitation and emission polarization dependent measurement of emitters was performed by rotating a half-wave plate (HWP) between the sample and a polarizer in the excitation and collection path, respectively. The polar plot of the polarization dependent emission intensity data was fitted by using $\cos^2(\theta)$ function, where $\theta/2$ is the HWP rotation angle.

For the second order correlation measurement, we placed a flip mirror in front of the spectrometer to guide the optical path to the Hanbury Brown and Twiss (HBT) interferometer. In this path, additional bandpass filters were used to spectrally purify the defect luminescence. The emitted light from *h*-BN defect was split by a 50:50 beam splitter and detected by two single-photon counting modules (SPCMs). Electrical signals from SPCMs were processed by a time-correlated single-photon counting (TCSPC) device to obtain the second order correlation data, $g^2(\tau)$, as a function of the time delay, $\tau$. $g^2(\tau)$ curve was fitted by $g^2(\tau) = 1 - \alpha e^{-|\tau/\tau_1|}$, where $\alpha$ is the fitting parameter ($0 \leq \alpha \leq 1$) and $\tau_1$ is the time scale related to the excited state lifetime.

Raman spectroscopy is performed to monitor the Fermi level tuning of graphene electrodes. A 532 nm cw laser was used as an excitation source with power below 200 μW. We focused on the graphene Raman G peak which shows the spectral shift and narrowing as a function of the carrier concentration to extract the graphene's Fermi level tuning while applying voltages [53].

### Fermi level tuning calculation

To calculate the Fermi level tuning of the graphene electrode ($\Delta E_F$) as a function of an applied voltage ($V$), we use the following relation [53,54]

$$eV = \Phi + 2\Delta E_F = e\left(\frac{en_{Gr}}{C_{Gr}}\right) + 2\Delta E_F$$

where $\Phi$ is the energy offset between the Dirac point of two graphene and $\Delta E_F$ is the Fermi level shift of the graphene at the Dirac point. The first term comes from the geometric capacitance of the graphene using $C_{Gr} = \frac{\varepsilon_0 \varepsilon_{BN}}{t_{BN}}$, where $\varepsilon_0$, $\varepsilon_{BN}$, and $t_{BN}$ are the vacuum permitivity, relative permitivity of $h$-BN, and the thickness of $h$-BN, respectively. The second term is attributed to the quantum capacitance of each graphene electrode. Since the charge concentration of graphene can be calculated by $n_{Gr} = \int_{min}^{max} \rho_{Gr}(E)dE$ where $\rho_{Gr}(E)$ is the density of state and $\rho_{Gr}(E) = \frac{E}{\pi \hbar^2 v_F^2}$, we get $n_{Gr} = \frac{\Delta E_F^2}{\pi \hbar^2 v_F^2}$. Here $\hbar$ and $v_F$ are the Plank constant and Fermi velocity, respectively. Finally the relation between the Fermi level tuning and the applied voltage is given by

$$eV = \frac{e^2}{\pi \hbar^2 v_F^2} \frac{t_{BN}}{\varepsilon_0 \varepsilon_{BN}} \Delta E_F^2 + 2\Delta E_F.$$

Applying the above relation to our experimental results, the voltage tuning range from -9 to 9 V yields the range of the Fermi level shift over the range of about 500 meV (See supplementary figure S5). This range is consistent with the independent measurement of the graphene's Fermi level tuning obtained by Raman spectroscopy (figure 2(e)) [41].

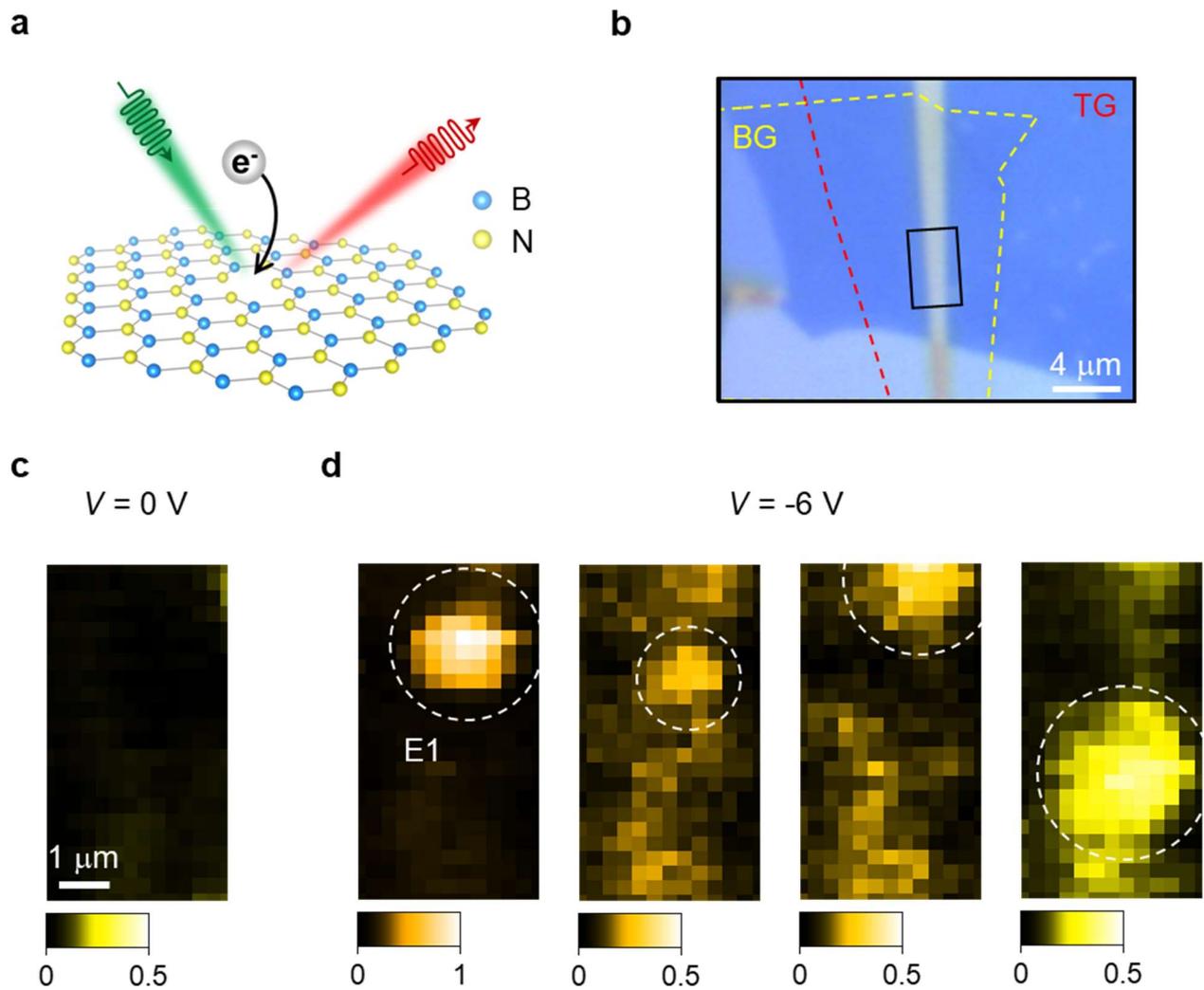

**Figure 1 Electrical switching of *h*-BN emitter photoluminescence.** (a) Schematics of the photoluminescence of electrically doped *h*-BN local defects. Green and red coloured rays indicate excitation laser and emitted light, respectively. B: boron. N: nitrogen. (b) Microscopic image of the device. The red and yellow dashed lines indicate the top (TG) and bottom (BG) graphene regions, respectively. The solid box indicates the region shown in (c-d). (c) 2D photoluminescence scan image of *h*-BN device measured in the absence of an applied voltage. Scale bar: 1 μm. (d) 2D scan image upon the application of an external voltage of -6 V, showing the localized photon emission. The positions of the localized emitters are indicated by white dashed lines. Emission intensity maps are shown for the wavelengths at 616, 609, 605, and 584 nm from the left to right panels, respectively.

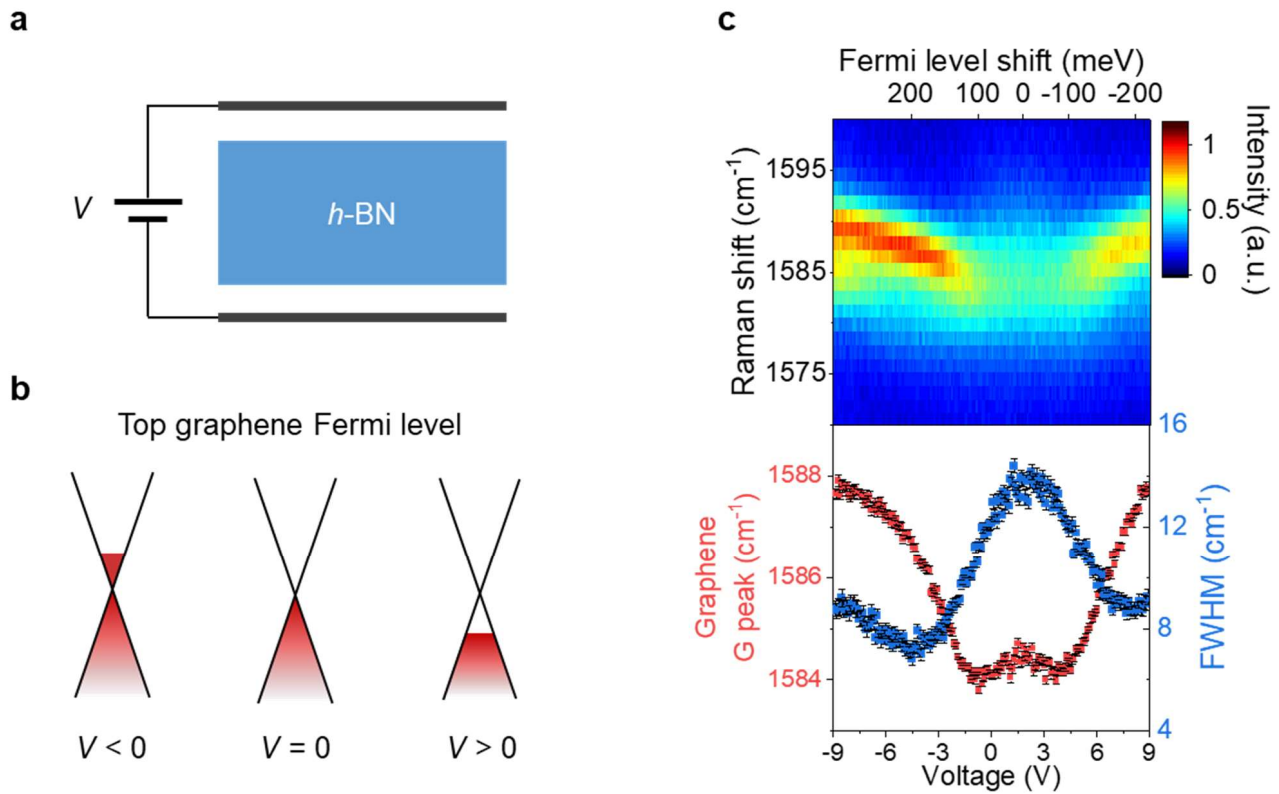

**Figure 2. Raman spectroscopy probing the Fermi level shift of graphene layer.** (a) Device schematics of graphene-stacked h-BN heterostructure. (b) Fermi level tuning of top graphene for negative (left), zero (centre), and positive (right) applied voltages. (c) (Upper) Spectra of graphene G Raman peak as a function of the applied voltage. (Lower) Raman centre peak position (red square) and full width at half maximum (FWHM) (blue square) obtained by Lorentzian fit functions. Fitting error bars are shown together.

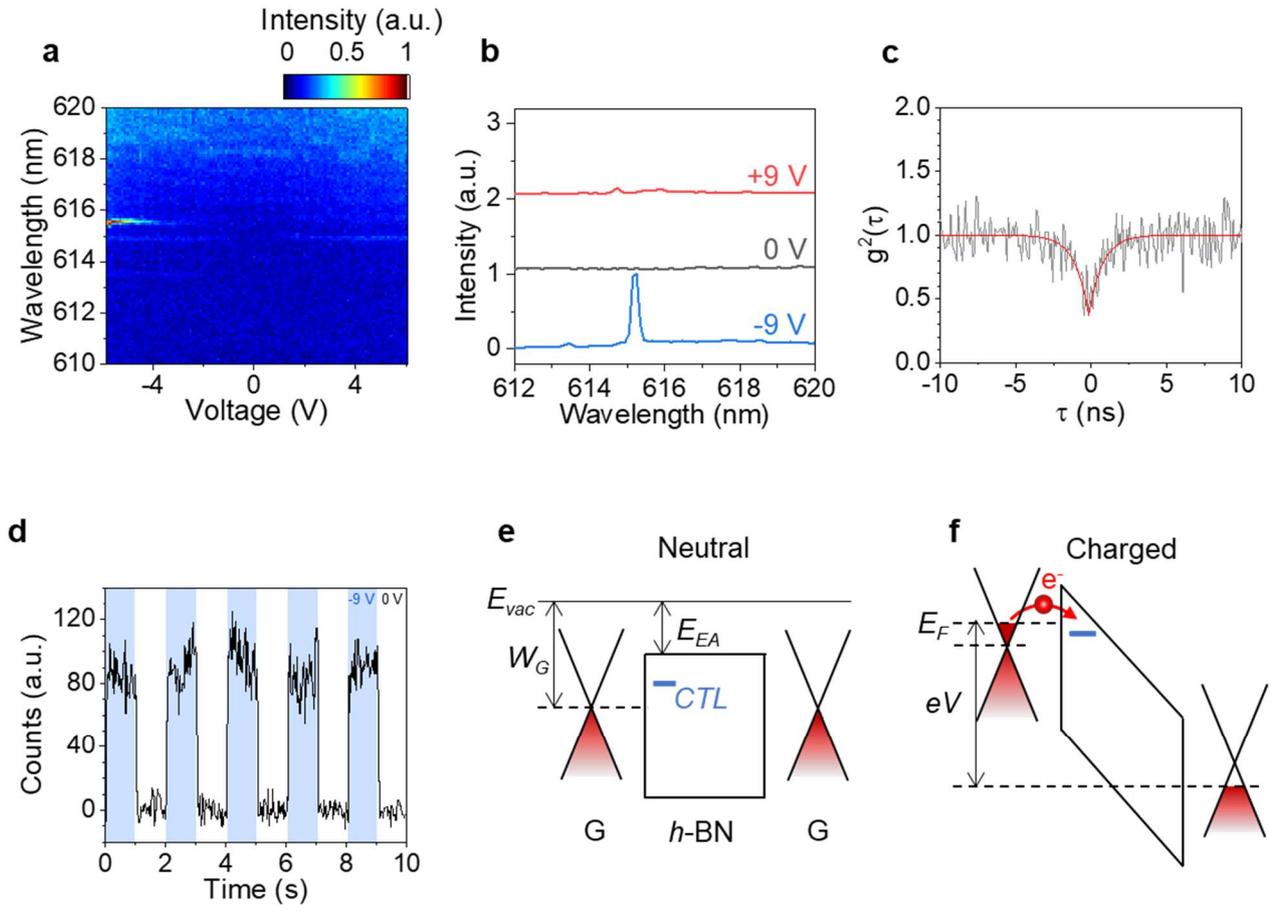

**Figure 3. Emitter with asymmetric voltage activation and the charge transition from nearby graphene.** (a) Colour plot of the voltage-activated emitter intensity as functions of the wavelength and applied voltage. (b) Emission spectrum at 9 V (red), 0 V (black), and -9 V (blue). Spectral lines are intentionally vertically shifted. (c) The second order correlation function measured at -9 V, showing that the emitter is a single photon source. The black line is the measurement data and the red line is the fitting result using the following parameters: $g^2(0) = 0.372$ and $\tau_1 = 898$ ps. (d) Repeatability measurement showing the modulation of the emission light intensity by using the square wave voltage. The voltage is periodically changed between -9 and 0 V with the period of 2 s. (e) Band diagram of the heterostructure in the absence of an external voltage. $E_{VAC}$: Vacuum energy level. $W_G$: Work function of neutral graphene (~ 4.5 eV). $E_{EA}$: Electron affinity of *h*-BN (~ 2.3 eV). CTL: Charge transition level of *h*-BN quantum defect. (f) Band diagram when an external voltage ($V$) charges the *h*-BN quantum defect. Electron is transferred from graphene to *h*-BN emitter.

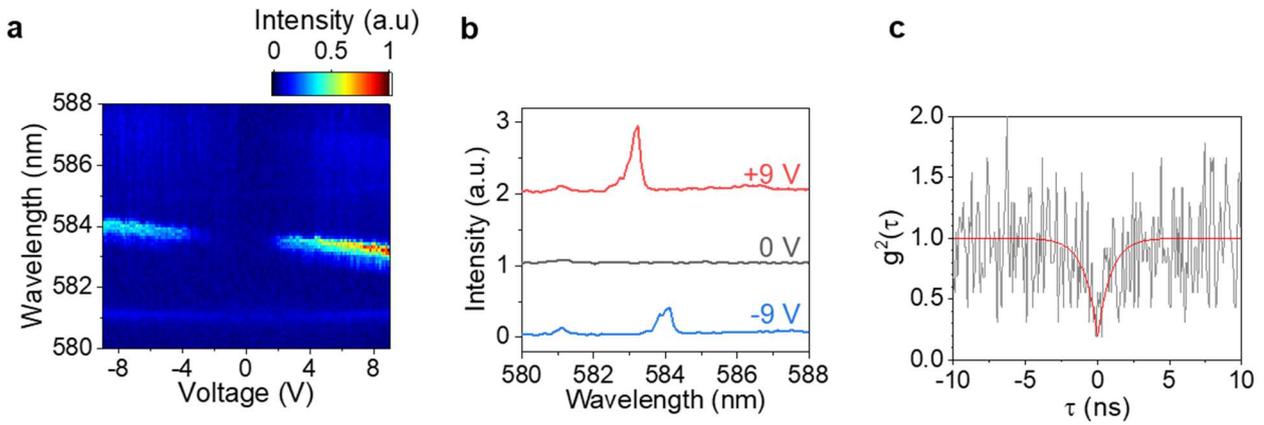

**Figure 4. Emitter with symmetric voltage activation.** (a) Colour plot of the voltage-activated emitter intensity as functions of the wavelength and applied voltage. (b) Emission spectrum at 9 V (red), 0 V (black), and -9 V (blue). Spectral lines are intentionally vertically shifted. (c) The second order correlation function measured at 7 V, showing that the emitter is a single photon source. The black line is the measurement data and the red line is the fitting result using the following parameters: $g^2(0) = 0.183$ and $\tau_1 = 900$ ps.

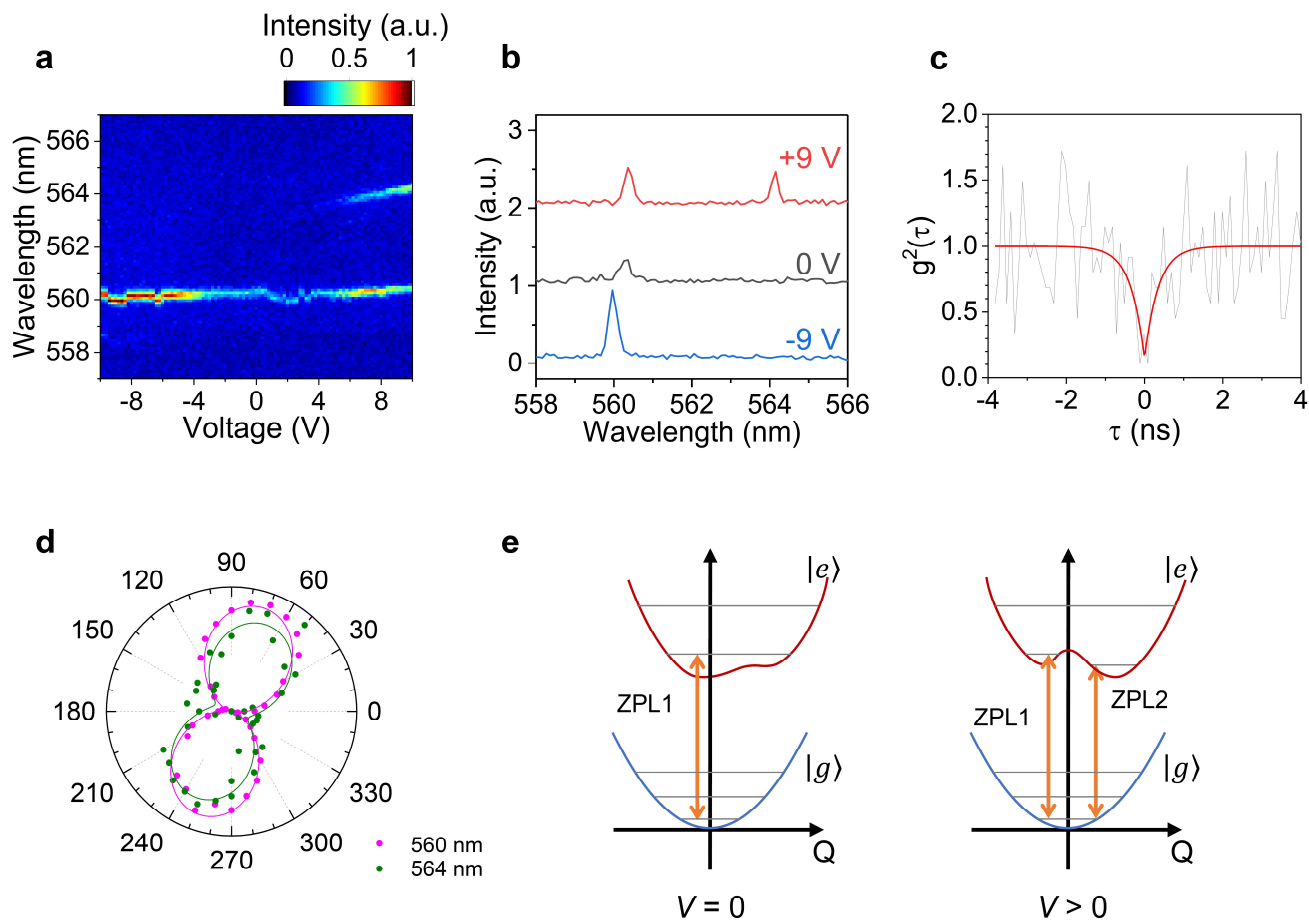

**Figure 5. Voltage-induced level splitting of *h*-BN emitter.** (a) Colour plot of the emitter peak intensities as functions of the wavelength and applied voltage. (b) Emission spectrum at 9 V (red), 0 V (black), and -9 V (blue). Spectral lines are intentionally vertically shifted. (c) The second order correlation function measured at -6 V, showing that the emitter is a single photon source. The black line is the measurement data and the red line is the fitting result using the following parameters: $g^2(0) = 0.150$ and $\tau_1 = 355$ ps. (d) Polar plots of the emission polarization dependence of the peak at 560 nm (magenta) and 564 nm (green), displaying almost the same polarization orientation. (e) Configurational coordinate diagrams depicting the evolution of the excited-state potential energy surface as a function of an external voltage. (Left) In the absence of an external voltage, one zero-phonon line (ZPL1) is possible, but there is a meta-stable point in the proximity of the excited-state zero-phonon vibrational state. (Right) Applying an external voltage beyond the threshold, a second potential well is formed, providing the second zero-phonon luminescence channel (ZPL2). *Q*: configurational coordinate. $|g\rangle$: ground state energy. $|e\rangle$: excited state energy.